\documentclass[graybox]{svmult}

\usepackage{mathptmx}       % selects Times Roman as basic font
\usepackage{helvet}         % selects Helvetica as sans-serif font
\usepackage{courier}        % selects Courier as typewriter font
\usepackage{type1cm}        % activate if the above 3 fonts are
\usepackage{makeidx}         % allows index generation
\usepackage{graphicx}        % standard LaTeX graphics tool
\usepackage{multicol}        % used for the two-column index
\usepackage[bottom]{footmisc}% places footnotes at page bottom
\usepackage{amssymb}
\usepackage{graphicx}
\usepackage{lipsum}

\newcommand\blfootnote[1]{%
  \begingroup
  \renewcommand\thefootnote{}\footnote{#1}%
  \addtocounter{footnote}{-1}%
  \endgroup
}

\makeindex             % used for the subject index
\begin{document}

\title*{Minimal Set of Texture Specific Quark Mass Matrices}
\author{Samandeep Sharma and Gulsheen Ahuja}
\institute{Samandeep Sharma \at Department of Physics, GGDSD College, Sector 32, Chandigarh \\
\email{sharma.saman87@gmail.com}
\and Gulsheen Ahuja \at Department of Physics, Panjab University, Chandigarh\\
\email{gulsheenahuja@yahoo.co.in}}
\maketitle
\blfootnote{**Based upon S. Sharma {\it et al.}  Phys.Rev. D91 (2015) 5, 053004.}
\abstract{Starting with the most general mass matrices, within the 
context of Standard Model and  some of its extensions, incorporating the ideas of weak basis transformations and `naturalness', we 
find that  there exists a particular set of texture specific quark mass matrices which can be considered to be the minimal
viable option. }

%\abstract{Each chapter should be preceded by an abstract (10--15 lines long) that summarizes the content. The abstract will appear \textit{online} at \url{www.SpringerLink.com} and be available with unrestricted access. This allows unregistered users to read the abstract as a teaser for the complete chapter. As a general rule the abstracts will not appear in the printed version of your book unless it is the style of your particular book or that of the series to which your book belongs.\newline\indent
%Please use the 'starred' version of the new Springer \texttt{abstract} command for typesetting the text of the online abstracts (cf. source file of this chapter template \texttt{abstract}) and include them with the source files of your manuscript. Use the plain \texttt{abstract} command if the abstract is also to appear in the printed version of the book.}

\section{Introduction}
\label{sec:1}
 One of the key challenges in the present day high energy physics
is to understand the vast spectrum of fermion masses and their
relationships with the corresponding mixing angles as well as mass
matrices. Despite impressive advances in the measurements of
fermion masses and mixing parameters, we are far from having a
compelling theory for flavor physics. Even for the case of quarks,
where precision measurements are available, the data is understood
in terms of phenomenological models having their roots in the
`bottom up' approach. In this context, exploring the possibility
of finding a minimal set of viable quark mass matrices can perhaps
be the first important step for solving the flavor riddle.

\par The bottom up approach of understanding fermion masses and
mixings has evolved in three different directions. Firstly, 
on the lines of Fritzsch ansatze \cite{frzans}, mass matrices are
formulated wherein certain elements of these are assumed to be
zero, usually referred to as texture zeros, and the compatibility
of the mixing matrix so obtained from these with the low energy
data ensures the viability of the formulated mass matrices. However,  despite
showing considerable promise, in this approach the
possibility to arrive at a minimal set of viable quark mass
matrices emerges only by carrying out an exhaustive case by case
analysis of all possible texture zero mass matrices \cite{singreview}.

\par Further, within the framework of SM
and its extensions, one has the freedom to make unitary
transformations, referred to as `Weak Basis (WB) transformations',
which change the mass matrices  without changing the quark mixing
matrix. Using WB transformations, several attempts
 \cite{fxwb}-\cite{costa} have been made wherein the above freedom
is exploited to introduce texture zeros in the quark mass
matrices.  This results in somewhat reducing the number of free
parameters of general mass matrices, however, in the absence of
any constraints on the elements of the mass matrices, leads to a
large number of texture zero matrices which are able to explain
the quark mixing data.

\par In yet another approach, advocated by Peccei and Wang \cite{nmm},
the concept of `natural mass matrices' has been introduced to
formulate viable set of mass matrices at the Grand Unified
Theories (GUTs) as well as the $M_Z$ scale. In order to avoid fine
tuning, the elements of the mass matrices are constrained in order
to reproduce the hierarchical nature of the quark mixing angles. This
results in constraining the parameter space available to the
elements of the mass matrices, however without yielding a finite
set of viable mass matrices at the GUTs as well as the $M_Z$
scale.

\par  A careful perusal of the above mentioned approaches suggests that
none of these leads to a finite set of viable texture specific mass
matrices, thus in order to obtain the same perhaps one needs
to combine the three.  The purpose of the present work is, therefore, to
follow the texture zero
approach coupled with WB transformations to reduce the number of
free parameters of general hermitian mass matrices as well as to
impose the condition of `naturalness' for constraining the
parameter space available to the elements of these.

\section{Methodology}
\label{sec:2}
To begin with, we consider the following hermitian mass matrices
\begin{equation}
 M_{q}=\left( \begin{array}{ccc}
E_q & A _{q} & F _{q}     \\ A_{q}^{*} & D_{q} &  B_{U}     \\
 F _{q}^{*} & B_{q}^{*}  &  C_{q} \end{array} \right) ~~~~~~~~~~ (q=U,D),
 \label{genmm}
\end{equation}
which, without loss of generality, are related to the most general
mass matrices \cite{singreview}. As a next step, one can introduce
texture zeros in these matrices using the WB transformations
\cite{fxwb}, in particular, one can find a matrix $W$ transforming
$M_U \rightarrow W^{\dagger} M_U W$ and $M_D \rightarrow
W^{\dagger} M_D W$, leading to
\begin{equation}
 M_{U}=\left( \begin{array}{ccc}
E_{U} & A _{U} & 0      \\ A_{U}^{*} & D_{U} &  B_{U}     \\
 0 &     B_{U}^{*}  &  C_{U} \end{array} \right), \qquad
M_{D}=\left( \begin{array}{ccc} E_{D} & A _{D} & 0\\ A_{D}^{*} & D_{D} &
B_{D}     \\
 0 &     B_{D}^{*}  &  C_{D} \end{array} \right).
\label{t20}\end{equation}
             The above matrices, wherein $A_{q}= |A_{q}|e^{i
\alpha_{q}}$ and $B_{q}=|B_{q}|e^{i \beta_{q}}$ for $q=U, D$, can
be characterized as texture 2 zero quark mass matrices.  Further, in
order to incorporate the condition
of `naturalness' on these mass matrices, we have considered the
following hierarchy for the elements of the matrices
\cite{unified}
\begin{equation}
(1,i) < (2,j) \lesssim (3,3);~~~~~~~~~~~~~~~~~i=1,~2,~3, ~~j=2,~3.
\label{nathier}
\end{equation}
 Therefore, the matrices given in equation
(\ref{t20}) can now be considered as most general and their
analysis can lead to very broad and interesting consequences.

Before getting into the details of the analysis, we first present
some of the essentials pertaining to the construction of the CKM
matrix from these mass matrices. Details in this regard can be looked up in 
\cite{singreview,uniqueprd}. To facilitate diagonalization,
for $q=U,D$, the mass matrix $M_{q}$ may be expressed as $ M_q =
Q_q^{\dagger} M_q^r Q_q $ implying $ M^{r}_q = Q_q M_q
Q_q^{\dagger}$ where $ M^{r}_q$ is a symmetric matrix with real
eigenvalues and $ Q_q $ is the diagonal phase matrix. The $M_{q}^r$ can be diagonalized using the following
transformations
\begin{equation}
M^{diag}_{q}= O^{T}_{q}M^{r}_{q}O_{q}= O^{T}_{q}Q_{q}
M_{q}Q^{\dagger}_q O_{q}= {\rm Diag}(m_{1}, -m_{2}, m_{3}), 
\end{equation}
where the subscripts 1, 2 and 3 refer respectively to $u$, $c$,
$t$ for the up sector and $d$, $s$, $b$ for the down sector. Using 
these diagonalizing transformations, the quark mixing matrix 
can be obtained from the relation

\begin{equation}
V_{\rm CKM} = O_{U}^{T} Q_{U} Q_{D}^{\dagger} O_{D}.
\label{ckmrel} 
\end{equation}
% Always give a unique label
% and use \ref{<label>} for cross-references
% and \cite{<label>} for bibliographic references
% use \sectionmark{}
% to alter or adjust the section heading in the running head

\section{Calculations and results}
To carry out the numerical work, the parameters $\phi_1$ and
$\phi_2$, related to the phases of the mass matrices, $\phi_1$ =
$\alpha_U- \alpha_D$ and $\phi_2= \beta_U- \beta_D$, have been
given full variation from 0 to $2\pi$. Apart from $\phi_1$ and
$\phi_2$, the free parameters $E_U$, $E_D$, $D_U$ and $D_D$ have
also been given wide variation in conformity with the condition of
naturalness. The
quark masses and the mass ratios at the $M_Z$ scale \cite{xing2012} have 
been used as inputs whereas the latest values \cite{pdg2014} of
CKM parameters, pertaining to three mixing angles and one CP violating
phase, have been used as constraints in our analysis.

Coming to the outcome of our analysis, using the relation between mass matrices
and mixing matrix, given in equation (\ref{ckmrel}), the resultant
CKM matrix comes out to be
 \begin{equation}
 {\rm V_ {CKM}}=\left( \begin{array}{ccc}
0.9739-0.9745 & 0.2246-0.2259 & 0.00337-0.00365  \\ 0.2224-0.2259
& 0.9730-0.9990 &  0.0408-0.0422    \\ 0.0076-0.0101 &
0.0408-0.0422 &  0.9990-0.9999 \end{array}
\right), \label{t20ckm}
\end{equation}
being fully compatible with the one given by PDG \cite{pdg2014}.
Also, the CP violating Jarlskog's rephasing invariant parameter
$J$ comes out to be $(2.494-3.365) \times 10^{-5}$ which again is
compatible with its latest experimental range, $(3.06
^{+0.21}_{-0.20})\times 10^{-5}$.

\begin{figure}[hbt]
\begin{minipage}{0.40\linewidth}   \centering
\includegraphics[width=1.0\textwidth,angle=-90]{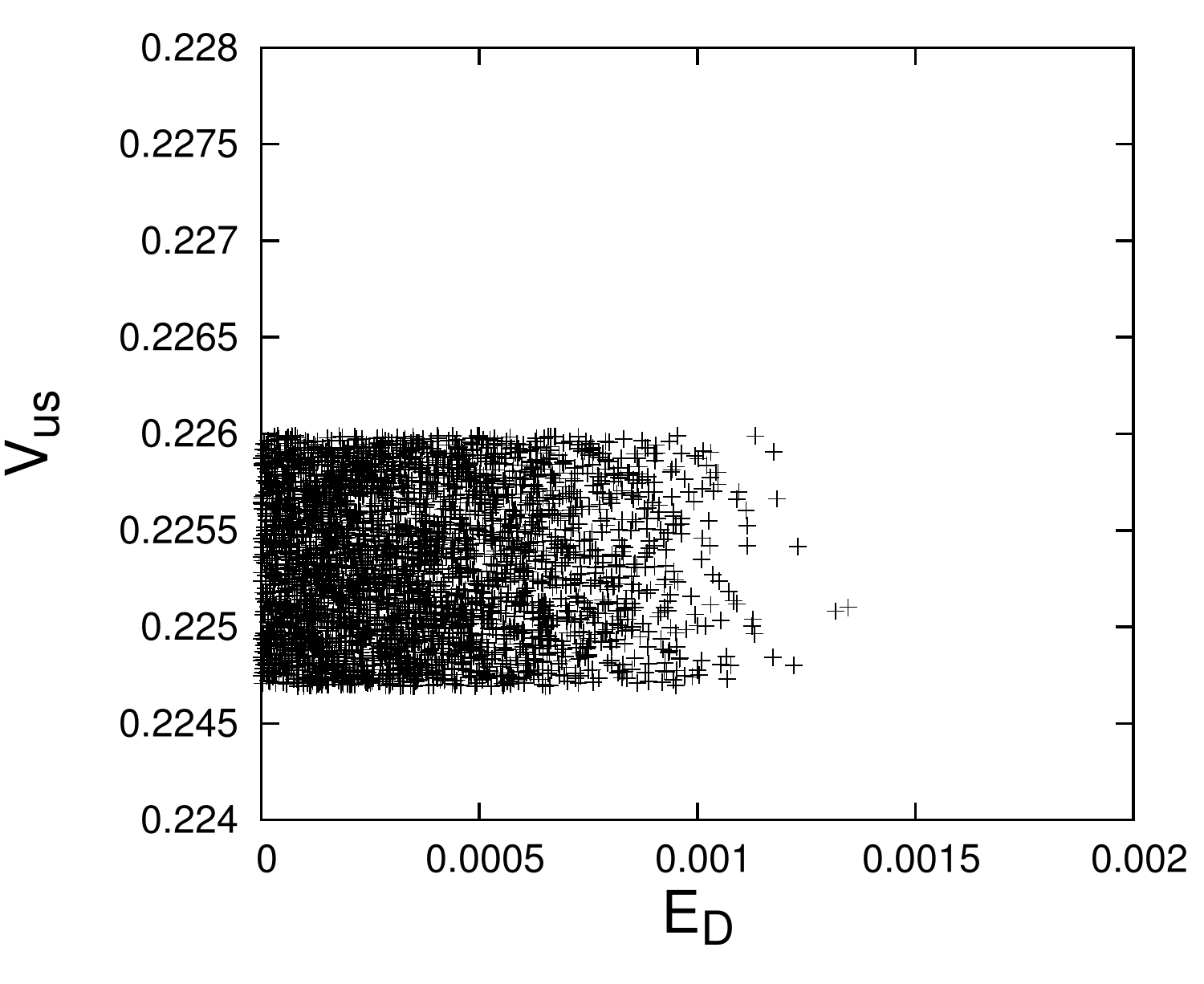}
\end{minipage} \hspace{1.5cm}
\begin{minipage} {0.40\linewidth} \centering
\includegraphics[width=1.0\textwidth,angle=-90]{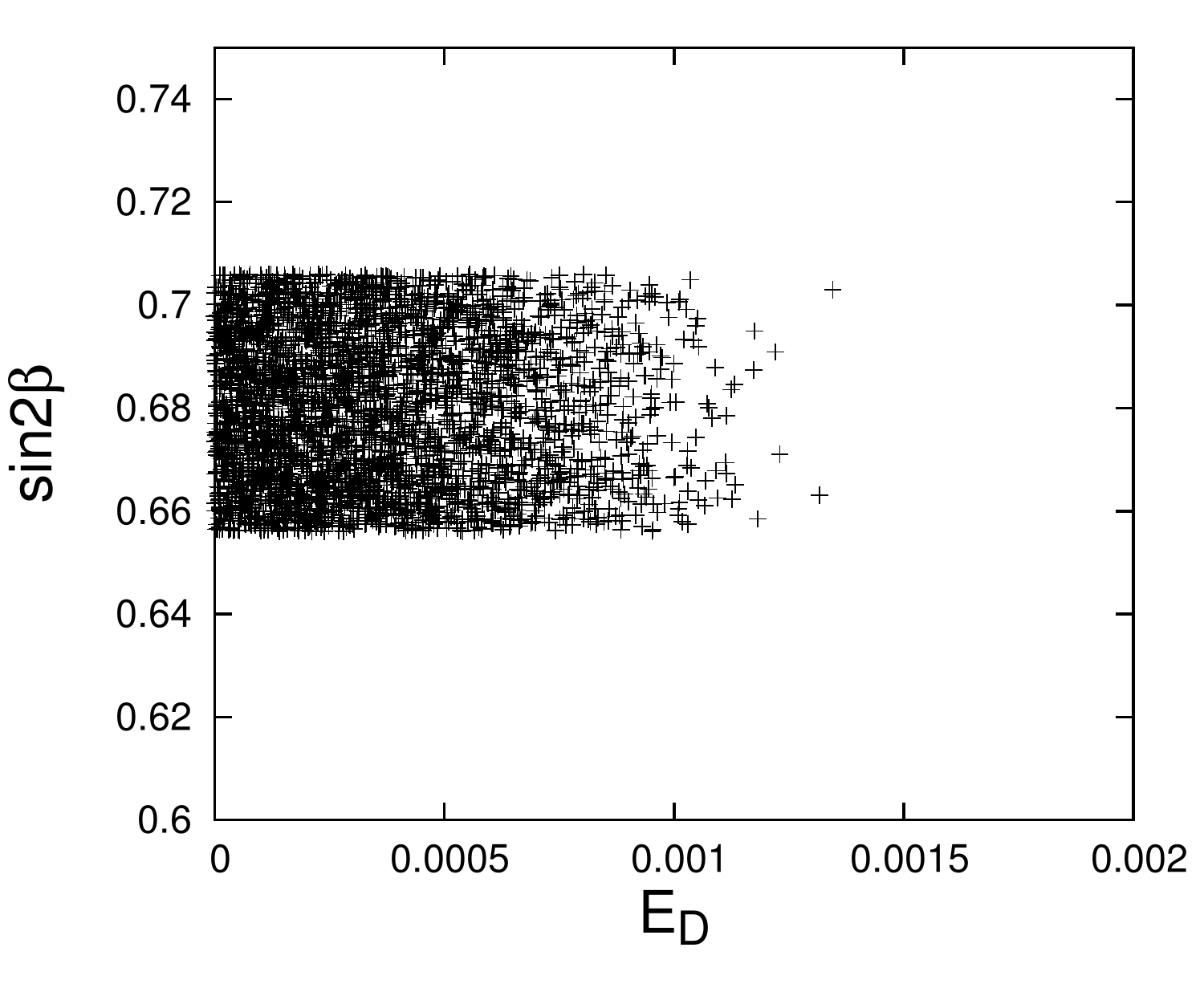}
 \end{minipage} \hspace{0.5cm}
\caption {Plots showing the dependence of $V_{us}$ and $\sin 2\beta$
on the parameter $E_D$.} \label{fig1}
\end{figure}

The viability of these general texture 2 zero mass matrices $M_U$
and $M_D$ is beyond doubt, however 
examining the parameter space available to various elements of these
matrices, one finds that the (1,1) element of both $M_U$ and 
$M_D$ is not only quite small in
comparison with the other non zero elements but is also 
essentially redundant. This fact can clearly be understood from 
a careful look at the Figure (\ref{fig1}), wherein 
the parameter $E_D$ has been plotted against 
 $|V_{us}|$ and the CP asymmetry parameter Sin2$\beta$. Similar conclusions can be drawn from
$E_D$ versus the other CKM matrix elements plots. In the up
sector, similar plots pertaining to (1,1) element $E_U$ of the
matrix $M_U$ reveal that again this parameter is also quite small
and essentially redundant.

\subsection{Texture four zero mass matrices}
Keeping in mind the above discussion, ignoring the elements $E_U$ and $E_D$ of the mass
matrices, one gets $M_U$ and $M_D$ as
\begin{equation}
 M_{U}=\left( \begin{array}{ccc}
0 & A _{U} & 0      \\
A_{U}^{*} & D_{U} &  B_{U}     \\
 0 &     B_{U}^{*}  &  C_{U} \end{array} \right), \qquad
M_{D}=\left( \begin{array}{ccc} 0 & A _{D} & 0\\
A_{D}^{*} & D_{D} & B_{D}     \\
 0 &     B_{D}^{*}  &  C_{D} \end{array} \right),
\label{flt40}
\end{equation}
 indicating a transition from texture 2 zero mass
matrices to texture 4 zero mass matrices. Carrying out a similar
analysis for these matrices, the corresponding CKM matrix comes
out to be
 \begin{equation}
 {\rm V_ {CKM}}=\left( \begin{array}{ccc}
0.9741-0.9744 & 0.2246-0.2259 & 0.00337-0.00365  \\ 0.2245-0.2258
& 0.9732-0.9736 &  0.0407-0.0422    \\ 0.0071-0.0100 &
0.0396-0.0417 &  0.9990-0.9992 \end{array} \right). \label{flckm}
\end{equation}
This
matrix is not only in complete agreement with the latest quark
mixing matrix given by PDG \cite{pdg2014}, but also is fully
compatible with the CKM matrix given in equation (\ref{t20ckm}).
Further, the range of the CP violating Jarlskog's rephasing
invariant parameter $J$ comes out to be $(2.50-3.37) \times
10^{-5}$ which again is compatible with its latest experimental
range, justifying our earlier conclusion that the elements $E_U$
and $E_D$ are essentially redundant as far as reproducing the CKM
parameters are concerned.

\par Further, it is interesting to note that 
using the WB transformations, apart from the matrices
given in equation (\ref{flt40}), one gets several other possible
texture 4 zero mass matrices which may or may not be related
through permutations. Based on whether the matrices are related
through permutations or not, all possible texture 4 zero mass
matrices can be classified as shown in Table (\ref{3t4}). The
matrices which are not related to each other through permutations
have been put into different categories.

\begin{table}
\caption{Table showing various
phenomenologically allowed texture 2 zero possibilities,
categorized into four distinct categories.}
\label{3t4}

 %\scalebox{0.6}{
\begin{tabular}{|c|c|c|c|c|c|c|} \hline
 & a & b  & c & d & e & f  \\
 \hline &&&&&& \\
Category 1 & $\left ( \begin{array}{ccc} {\bf 0} & A & {\bf 0} \\ A^{*}  & D
& B \\ {\bf 0} & B^{*}  & C \end{array} \right )$  &
 $\left ( \begin{array}{ccc} {\bf 0} & {\bf 0}  & A \\ {\bf
0}  & C & B \\  A^{*} & B^{*}  & D \end{array} \right )$  &   $\left (
\begin{array}{ccc} D & A &
  B\\ A^{*}  & {\bf 0}  & {\bf 0} \\  B^* & {\bf 0}
& C \end{array} \right )$ &  $\left ( \begin{array}{ccc} C & B & {\bf 0}
 \\ B^{*} & D & A  \\ {\bf 0} & A^{*} & {\bf 0}
 \end{array} \right )$ &  $\left ( \begin{array}{ccc} D & B & A
  \\ B^{*}  & C & {\bf 0} \\ A^{*} & {\bf 0}
& {\bf 0} \end{array} \right )$ & $\left ( \begin{array}{ccc} C & {\bf 0} & B
 \\ {\bf 0}  & {\bf 0}  & A \\B^* & A^*  &
D \end{array} \right )$ \\ \hline &&&&&&\\

Category 2 & $\left ( \begin{array}{ccc} D & A & {\bf 0} \\ A^{*}  & {\bf 0}
&  B \\ {\bf 0} & B^*  & C \end{array} \right )$     & $\left ( \begin{array}{ccc} D
& {\bf 0} & A
 \\ {\bf 0} & C & B \\  A^* & B^*  &
{\bf 0} \end{array} \right )$  & $\left ( \begin{array}{ccc} {\bf 0} & A &
 B \\ A^*  & D & {\bf 0}  \\  B & {\bf 0} &
C \end{array} \right )$ &  $\left ( \begin{array}{ccc} C & B & {\bf 0} \\ B^*  &
{\bf 0} &  A \\ {\bf 0} & A^*  & D \end{array} \right )$   &  $\left (
\begin{array}{ccc} C & {\bf 0} & B
 \\ {\bf 0} & D & A \\  B^* & A^*  &
{\bf 0} \end{array} \right )$
 & $\left ( \begin{array}{ccc} {\bf 0} & B &
 A \\ B^*  & C & {\bf 0}  \\  A^*  & {\bf 0} &
D \end{array} \right )$ \\ \hline &&&&&&\\

Category 3  &$\left ( \begin{array}{ccc} {\bf 0} & A &
 D\\ A^*  &{\bf 0}  & B \\ D^* & B^*  &
C \end{array} \right )$  &
  $\left ( \begin{array}{ccc} {\bf 0} &
D & A
\\  D^*  & C  & B \\  A^* & B  &
 {\bf 0}\end{array} \right )$ & $\left ( \begin{array}{ccc} {\bf 0} & A &
B
\\ A^* & {\bf 0} & D \\ B^*  & D^* &
C \end{array} \right )$ & $\left ( \begin{array}{ccc} {\bf 0} & B &
 C\\ B^*  &{\bf 0}  & A \\ C^* & A^*  &
D \end{array} \right )$   & $\left ( \begin{array}{ccc} {\bf 0} & C & B
\\  C^*  & D & A\\  B^* & A^*  &
 {\bf 0}\end{array} \right )$ & $\left ( \begin{array}{ccc} {\bf 0} & B &
A
\\ B^* & {\bf 0} & C \\ A^*  & C^*
& D \end{array} \right )$    \\ \hline &&&&&&\\

Category 4 & $\left ( \begin{array}{ccc} A & {\bf 0} & {\bf 0} \\ {\bf 0}  &
D & B\\ {\bf 0} & B^* & C \end{array} \right )$ & $\left (\begin{array}{ccc}C & {\bf
0} & B
 \\ {\bf 0}  & A & {\bf 0} \\ B^*  & {\bf 0}  &
D \end{array} \right )$ & $\left ( \begin{array}{ccc} C & B & {\bf 0} \\ B^*  & D &
{\bf 0} \\ {\bf 0} & {\bf 0}  & A \end{array} \right )$ &
    $\left ( \begin{array}{ccc} A & {\bf 0} & {\bf 0} \\
{\bf 0}  & C & B\\ {\bf 0} & B^* & D \end{array} \right )$ & $\left
(\begin{array}{ccc}D & {\bf 0} & B
 \\ {\bf 0}  & A & {\bf 0} \\ B^*  & {\bf 0}  &
C \end{array} \right )$  & $\left ( \begin{array}{ccc} C & B & {\bf 0} \\ B^*  & D &
{\bf 0} \\ {\bf 0} & {\bf 0}  & A \end{array} \right )$\\
  \hline
\end{tabular} 
%\vspace{0.7 cm}
%}  
\end{table}

Coming to the numerical analysis of the matrices listed in the
Table, for the matrices belonging to category 1, considering both
$M_U$ and $M_D$ as 1a type, corresponding to the ones mentioned in
equation (\ref{flt40}), we have already shown that these are
viable and explain the quark mixing data quite well. The other
matrices of this category, related through permutation matrix,
also yield similar results. For the matrices belonging to category
4, one finds that interestingly these are not viable as in all
these matrices one of the generations gets decoupled from the
other two. 
Further, for categories 2 and 3, again a similar
analysis reveals that the matrices of these classes are also not
viable as can be understood from the following CKM matrices
obtained for categories 2 and 3 respectively, e.g.,
\begin{eqnarray}
{\rm V_ {CKM}}=\left(\begin{array}{ccc} 0.9740-0.9744 & 0.2247-0.2260 &
0.0024-0.0099\\ 0.2205-0.2256 & 0.9509-0.9727 & 0.0596-0.2172 \\
0.0140-0.0445 & 0.0584-0.2127 & 0.9905-1.0000\end{array} \right),\\
\label{4bckm} 
%\end{equation}
%
%\begin{equation}
{\rm V_ {CKM}}=\left(\begin{array}{ccc} 0.9736-0.9744 &  0.2247-0.2260 &
0.0098-0.0331\\ 0.2226-0.2278 & 0.9549-0.9719 & 0.0659-0.1937\\
0.00007-0.0340 & 0.0694-0.1928 & 0.9810-0.9976\end{array}\right),
\label{5ackm} 
\end{eqnarray}
 these matrices showing no compatibility with the
latest quark mixing data. Therefore, the matrices pertaining 
to categories 2 and 3 can
be considered to be non viable.

\section{Summary and Conclusions}
To summarize, starting with the most general mass matrices, using
the concept of weak basis transformations, one first obtains
texture 2 zero quark mass matrices. Analysis of these matrices,
carried out by incorporating the naturalness condition, reveals
that certain elements are essentially redundant, therefore can be
discarded, reducing the matrices to texture 4 zero type.
Calculations pertaining to all the texture 4 zero mass matrices,
related through WB transformations, show that the mass matrices corresponding to 
the Fritzsch like texture four zero structure and its permutations can be 
considered to form a minimal set of fermion mass matrices in the quark sector. This
minimal set of texture structures for quarks
could be the first step towards unified textures for all fermions.

\begin{acknowledgement}
S.S.  would like to acknowledge Principal, GGDSD College, Sector 32, Chandigarh 
and the Chairperson, Department of Physics, P.U., for
providing facilities to work. 
\end{acknowledgement}
%
%%%%%%%%%%%%%%%%%%%%%%%% referenc.tex %%%%%%%%%%%%%%%%%%%%%%%%%%%%%%
% sample references
% %
% Use this file as a template for your own input.
%
%%%%%%%%%%%%%%%%%%%%%%%% Springer-Verlag %%%%%%%%%%%%%%%%%%%%%%%%%%
%
% BibTeX users please use
% \bibliographystyle{}
% \bibliography{}
%
\biblstarthook{
}

\end{document}